## Geophysical Research Letters

**Correspondence to:**
N. Østgaard,
Nikolai.Ostgaard@uib.no

# A new population of terrestrial gamma-ray flashes in the RHESSI data

N. Østgaard[1], K. H. Albrechtsen[1], T. Gjesteland[1,2], and A. Collier[3]

[1]Department of Physics and Technology, Birkeland Centre for Space Science, University of Bergen, Bergen, Norway,
[2]Department of Engineering Sciences, University of Agder, Grimstad, Norway, [3]School of Engineering, Electronic
Engineering Programme, University of KwaZulu-Natal, KwaZulu-Natal, South Africa

**Abstract** Terrestrial gamma-ray flashes (TGFs) are the most energetic photon phenomenon occurring
naturally on Earth. An outstanding question is as follows: Are these flashes just a rare exotic phenomenon
or are they an intrinsic part of lightning discharges and therefore occurring more frequently than previously
thought? All measurements of TGFs so far have been limited by the dynamic range and sensitivity of
spaceborne instruments. In this paper we show that there is a new population of weak TGFs that has not
been identified by search algorithms. We use the World Wide Lightning Location Network (WWLLN) to
identify lightning that occurred in 2006 and 2012 within the 800 km field of view of Reuven Ramaty High
Energy Solar Spectroscopic Imager (RHESSI). By superposing 740,210 100 ms RHESSI data intervals, centered
at the time of the WWLLN detected lightning, we identify at least 141 and probably as many as 191 weak
TGFs that were not part of the second RHESSI data catalogue. This supports the suggestion that the global
TGF production rate is larger than previously reported.

## 1. Introduction

With the discovery of terrestrial gamma-ray flashes (TGFs) [*Fishman et al.*, 1994] a new field of high-energy
atmospheric physics has emerged. As TGFs are the most energetic photon phenomenon occurring naturally
in the Earth's atmosphere, they may have an important impact on atmospheric electrodynamics. How severe
that impact is certainly depends on how common TGFs are. Are they just a rare exotic phenomenon or are
they an intrinsic part of lightning discharges? If the latter is the case, we have energy deposition both locally
and into the mesosphere and geospace that has not been accounted for.

When discovered, TGFs were considered to be a rare phenomenon. The Burst and Transient Source Exper-
iment, (BATSE) detected about 9 TGFs/yr (78 TGFs in 9 year) [*Nemiroff et al.*, 1997]. Although BATSE was an
instrument system with large detectors, the long trigger window (64 ms) lead to a strong bias toward only
the strongest TGFs. The Reuven Ramaty High Energy Solar Spectroscopic Imager (RHESSI) was a much smaller
instrument with an effective detection area of 256 cm$^2$, but as all data were downloaded, search algorithms
could be applied on ground to identify TGFs [*Smith et al.*, 2005]. The first RHESSI catalogue, which was based on
a rather conservative search algorithm, reported about 160 TGFs/yr (591 TGFs in 44 months) [*Grefenstette et al.*,
2009]. When a more sophisticated search algorithm was applied to the RHESSI data [*Gjesteland et al.*, 2012], the
number of identified TGFs/yr was more than doubled (∼350 TGF/yr). As all the new TGFs were weaker ones,
it is not unlikely that there exists a large population of weaker TGFs that we are not able to identify [*Østgaard
et al.*, 2012], either because they are intrinsically weak, produced at lower altitudes, or farther away from the
satellite's foot point. The next instrument detecting TGFs was Fermi Gamma Burst Monitor (GBM), which has a
slightly larger detector area of 320 cm$^2$. Fermi GBM was first using the low-energy detector (NaI) to trigger on
TGFs but increased the detection rate by a factor of 10 when using the BGO detectors for triggering [*Fishman
et al.*, 2011]. Fermi GBM is now using an optimized search scheme to identify TGFs and identifies about
850 TGFs/yr [*Briggs et al.*, 2013]. The Astririvelatore Gamma a Immagini Leggero (AGILE), due to its low incli-
nation (2.5°) and an onboard veto system, reported significantly less TGFs/yr [*Marisaldi et al.*, 2010]. However,
after disabling the veto system, AGILE has increased its detection rate of TGFs by a factor of 10 [*Marisaldi
et al.*, 2015].

All previous estimates of global production rate are based on the observed TGF rate and therefore limited
by instrument sensitivity. They were also affected by what is considered to be the effective detection field of
view. Estimates of global production rate of TGFs based on the first RHESSI catalogue ranged from 50 TGFs/d

[*Smith et al.*, 2005] using 1000 km field of view as effective detection area to 500 TGFs/d [*Carlson et al.*, 2009] using 300 km as effective area. Based on the number of TGFs measured by Fermi GBM the estimate of global production rate varies from 1100 TGFs/d [*Briggs et al.*, 2013] to 1200 TGFs/d [*Tierney et al.*, 2013]. Both studies accounted for the variable detection efficiency as a function of distance from subsatellite point.

By improving operational schemes and search algorithms, both the detection rate and estimated global production rate have increased. This is true for RHESSI, Fermi, and AGILE. The Airborne Detector for Energetic Lightning Emissions (ADELE) passed within 10 km (and 4 km) horizontal distance of 1213 (133) lightning discharges and detected only one (none) TGF and suggested an upper limit for global production rate of ~15,000 TGF/d. *Østgaard et al.* [2012] used the detection rates by RHESSI (first catalogue) and Fermi (first years) combined with their different sensitivity and orbits to estimate the most likely "true" fluence distribution of TGFs as measured from space. They found a power law distribution with an exponent of −2.3, which was later supported by dead-time-corrected data from Fermi (−2.2) [*Tierney et al.*, 2013] and AGILE (−2.4) [*Marisaldi et al.*, 2014]. *Østgaard et al.* [2012] also corrected the RHESSI TGF data for dead time effects and found the same power law. However, the corrected distribution had a signature of a roll-off at lower fluence, and two exponents were found to fit the data, −1.7 at the lower end and −2.6 at the higher end. Considering a power law with exponent of −2.3, this distribution could extend down to a sharp cut-off at ~1/100 of RHESSI lower threshold and still be consistent with the ADELE results. The global production rate would then be about 50,000 TGFs/d. If a roll-off with exponent of −1.7 was considered and still consistent with the ADELE results, they found that the hypothesis that all lightning produce TGFs cannot be excluded. Only future experiments with larger detectors from space or detectors flown closer to the lightning discharges can resolve whether there exists a large population of weaker TGFs that cannot be detected by the detectors flown so far. However, we can still explore whether there exists even weaker TGFs in existing data sets, which we are unable to identify with search algorithms. This is what we will report in this paper. A predecessor work to what we report here was presented by *Smith et al.* [2014].

## 2. Data and Method

Comparing TGFs and radio waves from lightning dates back to the time when TGFs were discovered [*Fishman et al.*, 1994] and a time coincidence of ±1.5 ms was reported [*Inan et al.*, 1996]. With the more accurate timing of the Fermi GBM, *Connaughton et al.* [2010] found matches between TGFs and WWLLN lightning to be within 40 μs. *Cummer et al.* [2011] reported coincidence on microsecond timescale between TGFs and low-frequency (LF) radio waves. Two TGFs detected by Fermi were related to lightning discharges detected by the National Lightning Detection Network (NLDN). Given the uncertainties of both instruments, the time coincidence was determined down to ±18 μs accuracy. This close time coincidence was further supported by *Connaughton et al.* [2013] who defined simultaneous Fermi GBM TGFs and WWLLN lighting to be within ±200 μs. They showed that the number of simultaneous TGFs and VLF signals increased when the duration of the TGFs decreased, consistent with the current pulse from the TGF being larger for shorter TGFs. These studies indicate that there might be many more TGFs simultaneous with the WWLLN signals. Whether this is because the TGF produces the radio signal or there are processes occurring simultaneously is not the focus of this paper. We only use the fact that WWLLN signals and TGFs occur simultaneously on microsecond timescale.

The analysis is based on data from 2006 and 2012, as only these 2 years of WWLLN data were available for us at the time of the analysis. We have used lightning located by WWLLN within a radius of 800 km from the foot point of RHESSI. This is a compromise between a large radius (1000 km) that would give us many background counts and a small radius (400 km) that would exclude many weak TGFs. Taking into account the systematic delay of the RHESSI clock of 1.8 ms [*Østgaard et al.*, 2013] and the propagation time from source location (WWLLN), we can align the RHESSI data with the WWLLN timing.

The timing uncertainty of WWLLN lightning depends on location and has been reported to range from 30 μs [*Connaughton et al.*, 2010] to 45 μs [*Østgaard et al.*, 2013], while *Hutchins et al.* [2012] reported <15 μs for 54% of the sferics. The uncertainty of the RHESSI clock is about ~100 μs, which is based on matches between RHESSI TGFs from the second catalogue [*Gjesteland et al.*, 2012] and WWLLN. Both these uncertainties are smaller than the average duration of TGFs of ~200–250 μs [*Gjesteland et al.*, 2010; *Fishman et al.*, 2011], and we will bin the data in 300 μs bins.

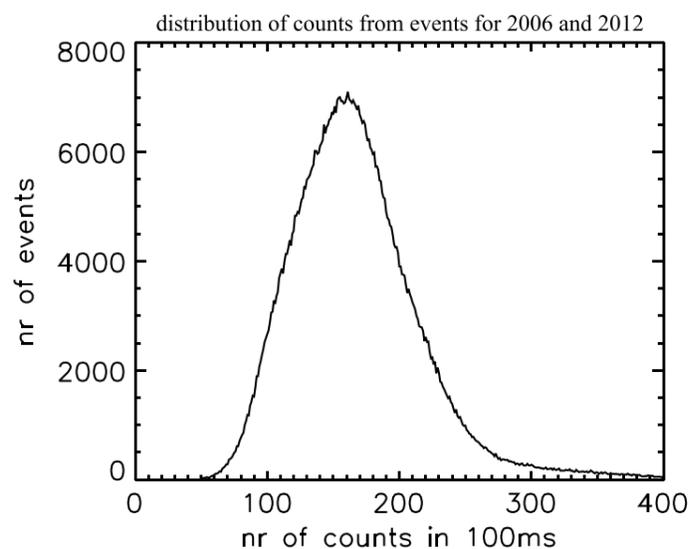

**Figure 1.** The distribution of background counts in 100 ms bins between 50 and 400.

All the 85 TGFs with WWLLN matches from 2006 and 2012 already identified by the search algorithm of the second RHESSI catalogue [*Gjesteland et al.*, 2012] are excluded from the analysis.

## 3. Results

For this analysis we have identified WWLLN lightning from 2006 and 2012 located within the radius of 800 km from RHESSI foot point. To avoid RHESSI data with very small background values (which are probably due to instrumental effects) and very high background values (which could be due to radiation belts, South Atlantic Anomaly or instrumental effects), we only use RHESSI data when the background count is between 50 and 400 in a 100 ms time interval. This leaves us with 740,210 lightning events. The distribution of background counts per 100 ms for these 740,210 RHESSI events are shown in Figure 1. The average of this distribution is 168 counts, which gives an average of 0.50 counts per 300 μs.

In Figure 2 we show the superposed 740,210 intervals of 80 ms of RHESSI data centered at the time of the WWLLN lightning. The average background count is 370,212 per 300 μs which gives an average of 0.50 counts per 300 μs for each RHESSI data string. The standard deviation, σ, is 608 counts. The peak at 0.00 s is 2903 counts above the average or 4.77σ and indicates hundreds of weak TGFs.

To be able to separate these TGFs from the background distribution, we want to examine whether the background in Figure 1 follows a Poisson distribution, given as follows:

$$F(k) = A \frac{\alpha^k}{k!} e^{-\alpha} \tag{1}$$

where $k$ is the number of counts per 300 μs, $A$ is the total number of events (740,210), and $\alpha$ is the average per 300 μs (0.50 counts).

Figure 3 shows the distribution of background counts per 300 μs in black overlaid the expected Poisson distribution (equation (1)) in red. These are the counts from Figure 2 when the center bin ±5 bins are excluded. As can be seen, the observed background follows a Poisson distribution, but there are deviations from 3 to 7 counts, which means that there are other variations in the background. This could also be seen from the width of the background distribution in Figure 1. A true Poisson distribution should be narrower with a standard deviation of 13 counts ($\sqrt{168}$), while we observe a standard deviation closer to 40 counts.

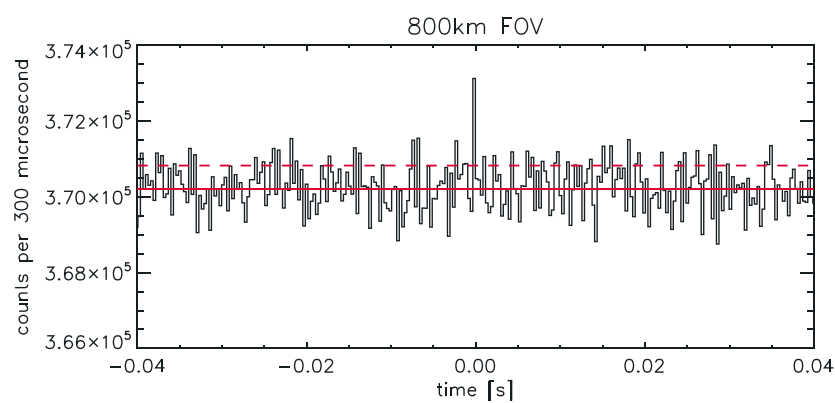

**Figure 2.** Superposed RHESSI 80 ms of data for 740,210 WWLLN lightning events found within 800 km radius of RHESSI foot point. RHESSI time is backpropagated to source location, and center time is the WWLLN time. The 85 known TGFs found by the search algorithm (second catalogue) are excluded. Vertical line is the average background counts, and dashed line is the standard deviation (σ). Bin size is 300 μs.

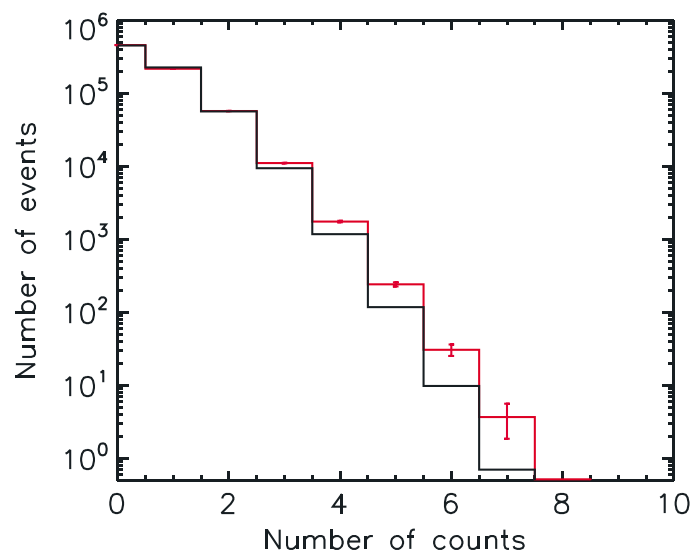

**Figure 3.** The distribution of background counts per 300 μs in red with the expected Poisson distribution in black.

The broader distribution is probably due to varying background at different latitudes [*Smith et al.*, 2002] resulting in a composition of many Poisson distributions.

In Figure 4 we show the distribution of counts in the center bin with the distribution of background counts from Figure 3 shown in red. Here we can clearly identify a new population of TGFs from 8 counts and above. At 7 counts we have a total of 48 on top of a background of 4. Even at 6 counts we have 81 total with a background of 31 and we should be able to separate the ~50 TGFs by examining each of the 81 more carefully. We have identified 141 new TGFs (7 or more counts) and probably 50 more if we are able to separate the TGF events with 6 counts from background of 6 counts. Compared to the 85 TGFs with WWLLN matches in the second catalogue (which we have excluded), we have almost tripled the number of WWLLN matches in the RHESSI data (factor of 2.7 – 3.2) from 2006 and 2012.

If we assume that the count distribution in the center bin is a combination of a Poisson distributed background and a power law distribution of weak TGFs, we can fit the distribution with the following equation:

$$F(k) = A \frac{\alpha^k}{k!} e^{-\alpha} + Bk^{-\lambda} \tag{2}$$

where $k$ is the number of counts pr 300 μs, while $A$, $B$, $\alpha$, and $\lambda$ are kept as free parameters. As we only have one event in the thirteenth, fifteenth, and seventeenth bins, we only use count bins from 0 to 12 for this fit. The best fit solution is shown by the black curve and the parameters are the following: $A = 737,056$ (which is fairly close to the 740,210 lightning we have), $\alpha = 0.50$ which is exactly the average background counts in a 300 μs bin, $B = 1311$, and $\lambda = 1.85$. Although care should be taken when comparing a power law distribution based on TGFs that have not been corrected for dead time and are obtained later in the RHESSI mission when there has been a degradation of the instrument, it is interesting to notice that this power index of 1.85 is very close to the roll-off index of 1.7 suggested by *Østgaard et al.* [2012]. If we had included the 85 WWLLN matches from the second catalogue, the power index becomes 1.95.

## 4. Discussion and Summary

We have identified 740,210 WWLLN lightning events from 2006 and 2012, within a radius of 800 km from RHESSI's foot point. By taking into account the systematic delay of the RHESSI clock and propagating the RHESSI data back to the time at the source location, we have identified a population of 141–191 weak TGFs from 2006 and 2012 that were not identified by our search algorithm. If we assume that 75% of the total lightning are intracloud (IC) lightning [*Boccippio et al.*, 2001] and that WWLLN is twice as sensitive to cloud-to-ground (CG) than IC lightning [*Abarca et al.*, 2010], we can expect that ~60% of the WWLLN lightning to be IC lightning. This means that of the 444,000 IC lighting in this data set, we have identified a total of 226–276 TGFs (new and old). This does not increase the TGF/IC-lighting ratio (1/1900–1/1600) or the global production rate (1300–1600 TGFs/d) very much, but it supports the idea that there may exist a large population of TGFs which is too weak to be observed by current instruments, either because they are farther away or they are produced at lower altitudes. Our choice of identifying lightning within a radius of 800 km from the RHESSI's foot point was essential as >1/2 of the new TGF population was found > 400 km from the foot point (not shown).

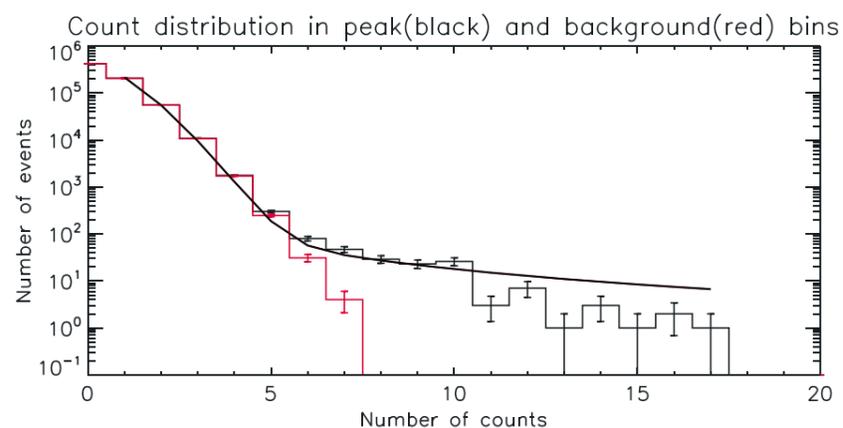

**Figure 4.** The distribution of counts in the center bin shown in black. The background distribution from Figure 3 is again shown in red. The black curve is a fit to the center bin distribution from 0 to 12 counts using equation (2).

Our results also give some support to the idea that the fluence distribution of TGFs has a roll-off at the lower end. In a recent paper, *McTague et al.* [2015] did a similar study based on Fermi GBM TGFs and lightning detected by NLDN and did not find any weak TGFs that were not already identified by the Fermi GBM search method [*Briggs et al.*, 2013]. Their no result can probably be explained by the statistics. The study was based on 1787 IC lightning, and if a similar identification ratio as we report (1/1600–1/1900) applies to their search, the probability of seeing none when ~1 is expected is rather big. Our study indicates that it may be possible to identify a new population of weaker TGFs by using lightning detection in the field of view of Fermi GBM and AGILE as well.

**Acknowledgments**
This study was supported by the European Research Council under the European Unions Seventh Framework Programme (FP7/2007-2013)/ERC grant agreement 320839 and the Research Council of Norway under contracts 208028/F50, 216872/F50 and 223252/F50 (CoE). We thank the RHESSI team and D.M. Smith for the use of RHESSI raw data and software. We thank the institutions contributing to WWLLN (http://wwlln.net/).